\documentstyle [12pt]{article}
\title{{\bf Study on the rare radiative decay \\   
$B_c \rightarrow D_s^*\gamma$ in the standard model and multiscale walking 
technicolor model }}
\author{Gongru Lu$^{a,b}$, Chongxing Yue$^{a,b}$, Yigang Cao$^a$, 
Zhaohua Xiong$^a$, Zhenjun Xiao$^{a,b}$\\
\small a. Physics Department of Henan Normal University, Xinxiang, Henan, 
453002, P. R. China$^*$\\
\small b. CCAST (world laboratory), P.O.Box 8732, Beijing, 100080, P. R. 
China\\}
\date{}

\begin{document}
\maketitle
\begin{abstract}	
Applying the perturbative QCD ( PQCD ) method, we study the decay 
$B_c\rightarrow D_s^*\gamma$ in the standard model and multiscale 
walking technicolor model. In the SM, we 
find that the contribution of weak annihilation is more important than that 
of the electromagnetic penguin. The presence of Pseudo-Goldstone-Bosons ( 
PGBs ) in MWTCM leads to a large enhancement in the rate of $B_c\rightarrow 
D_s^*\gamma$, but this model is in conflict with the branching ratio of
$Z\rightarrow b\overline b$ ( $R_b$ ) and the CLEO data on the branching 
ratio BR ( $b\rightarrow s\gamma$ ). If topcolor is 
further introduced, the calculated results in the topcolor assisted MWTCM 
can be suppressed and be in agreement with the CLEO data for a certain 
range of the parameters. 
\end{abstract}

\vspace{0.5cm}


\vspace{0.5cm}

* mailing address

\vspace{1CM}
\begin{flushleft}    
\section*{1. Introduction}
\end{flushleft}

The inclusive rare decay $B\rightarrow X_s \gamma$ has been studied several 
years ago [1]. Recently the physics of $B_c$ meson has caught 
intensive attentions[2]. The $B_c$ meson is believed to be the next and 
the final family of $B$ mesons, it provides unique opportunity to examine 
various heavy quark fragmentation models, heavy quark spin-flavor symmetry, 
different quarkonium bound state models and properties of inclusive decay 
channels. Furthermore, the radiative weak decays of $B_c$ meson also offer 
a rich source to measure CKM matrix elements 
of the standard model ( SM ). In this paper, we will address $B_c$ 
radiative decay $B_c\rightarrow D_s^*\gamma$.

Different from the decay $B\rightarrow X_s\gamma$ which is mainly induced by 
the flavor-changing $b\rightarrow s\gamma$ neutral currents [3], the bound 
state effects in the decay $B_c\rightarrow D_s^*\gamma$ may be rather large. 
Bound state effects include modifications from weak annihilation which 
involve no neutral flavor-changing currents at all. The effects of weak 
annihilation mechanism are expected rather large due to the large CKM 
amplitude. We will address this point in detail below.

Unfortunately, the well-known chiral-symmetry [4] and the heavy quark 
symmetry [5] can not be applied to this process. Recently, a perturbative 
QCD ( PQCD ) analysis of $B$ meson decays seems give a good prediction [6]. 
As it is argued in Ref.[7] that two body nonleptonic decay of $B_c$ meson 
can be conveniently studied within the framework of PQCD suggested by 
Brodsky-Lepage [8] and then developed in Ref.[6]. Here, we preview the 
reliability of PQCD analysis of $B_c$ radiative decay: in the process $b
\rightarrow s\gamma$, s quark obtains large momentum by recoiling, in order 
to form a bound state with the spectator $\overline c$ quark, the most 
momentum of s quark must be transferred to $\overline c$ by a hard scattering 
process. PQCD [6, 8] can be used in the calculation for the hard scattering 
process because the heavy charm usually share the most momentum of final 
state ( i.e. $D_s^*$ ). The relevant Feynman diagrams are given in Fig.1. 

Like in $B\rightarrow K^*\gamma$, the subprocess $b\rightarrow s\gamma$ in 
$B_c\rightarrow D_s^*\gamma$, is usually controlled by the one-loop 
electromagnetic penguin diagrams ( Fig.1.a ). It plays an important role in 
testing loop effects in the SM and in searching for the 
physics beyond the SM ( so called new physics ).

Most recently, the contribution of the electromagnetic penguin interaction 
to the branching ratio BR ( $b\rightarrow s\gamma$ ) from PGBs in the one 
generation technicolor model ( OGTM ) has been estimated in Ref.[9]. However, 
we know that there are some problems ( such as flavor-changing neutral 
currents ( FCNCs ), the large positive contributions to the parameters ) in 
most conventional TC models. Walking technicolor ( WTC ) has been advocated 
as a solution to the problem of large flavor-changing neutral current 
interactions in extended technicolor ( ETC ) theories of quark and lepton 
mass generation [10]. Furthermore, the electroweak parameter $S$ in WTC 
models is smaller than that in the simple QCD-like ETC models and 
consequently its deviation from the SM value may fall within current 
experimental bounds [11]. To explain the large hierarchy of the quark masses, 
multiscale WTC models ( MWTCM ) are further proposed [12].

However, as discussed in Ref.[13], the correction of PGBs in MWTCM to the 
$Z\rightarrow b\overline b$ branching ratio ( $R_b$ ) is too large when 
compared with recent LEP data. In this paper we calculated the contribution 
to the branching ratio $B_c\rightarrow D_s^*\gamma$ from the PGBs in MWTCM 
and found that such contribution is too large when compared with CLEO 
constraint for the inclusive decay $b\rightarrow s\gamma$. In general, there 
are two mechanisms which contribute to the decay $B_c\rightarrow D_s^*
\gamma$: one proceeds through the short distance $b\rightarrow s\gamma$ 
transition while the other through weak annihilation accompanied by a photon 
emission. On the other hand, if topcolor [14] is further introduced to the 
multiscale walking technicolor model, the modification from the PGBs in 
the topcolor assisted multiscale walking technicolor model ( TAMWTCM ) to 
$B_c\rightarrow D_s^*\gamma$ is strongly suppressed,  and therefore can be 
consistent with the recent CLEO data for the branching ratio BR ( 
$b\rightarrow s\gamma$ ) [15]. 
	
This paper is organized as follows: In Sec.2, we display our calculations in 
the SM and MWTCM. We present the final numerical results in Sec.3. Sec.4 
contains the discussion.
\begin{flushleft}
\section*{2. Calculation}
\end{flushleft}

Using the factorization scheme [8] within PQCD, the momentum of quarks
are taken as some fractions x of the total momentum of the meson weighted by
 a soft physics distribution functions $ \Phi_{H}(x) $. The peaking 
approximation is used for $\Phi_H(x)$ [16], the distributition amplitude of 
$B_c$ and $D_s^{*}$ are
$$
\Phi_{B_c}(x)=\frac{f_{B_c}\delta(x-\epsilon_{B_c})}{2\sqrt{3}},
\eqno{(1.a)}
$$
$$
\Phi_{D_s^{*}}(x)=\frac{f_{D_s^{*}}\delta(x-\epsilon_{D_s^{*}})}{2\sqrt{3}}, 
\eqno{(1.b)}
$$
where $f_{B_c}$, $f_{D_s^*}$ are decay constants of $B_c$ and $D_s^*$ 
respectively, and                
$$
\epsilon_{B_c}=\frac{m_c}{m_{B_c}},                  
\eqno{(1.c)}
$$
$$
\epsilon_{D_s^*}=\frac{m_{D_s^*}-m_c}{m_{D_s^*}}.
\eqno{(1.d)}
$$
The spinor parts of $B_c$ and $D_s^{*}$ wave functions are
$$
\frac{(\not p+m_{B_c})\gamma_5}{\sqrt{2}}, 
\eqno{(2.a)}
$$
$$
\frac{(\not p-m_{D_s^*})\not \epsilon}{\sqrt{2}},         
\eqno{(2.b)}
$$
where $\epsilon $ is the polarization vector of $D_s^*$.

\begin{flushleft}
\subsection*{2.1. Electromagnetic penguin contribution}
\end{flushleft}

The short distance electromagnetic penguin process is governed by the 
electromagnetic penguin operators [1]. At the 
weak scale $\mu=m_b$, the effective Hamiltonian for $b\rightarrow s\gamma$ 
transition is 
$$		
H_{eff}=\frac{4G_F}{\sqrt{2}}V_{tb}V_{ts}^* C_7(m_b)O_7,
\eqno{(3)}
$$
where 
$$
O_7=\frac{em_b{\overline s}\sigma_{\mu\nu}F^{\mu\nu}(1+\gamma_5)b}{32\pi^2}
\eqno{(4.a)}
$$	
and which is denoted by a blob in Fig.1.a. The corresponding coefficient of 
$O_7$ has the form 
$$
C_7(m_b) =\varrho^{-\frac{16}{23}}[C_7(m_W)+\frac{8}{3}
(\varrho^\frac{2}{23}-1)C_8(m_W)]+C_2(m_W)\sum\limits_{i=1}\limits^{8}h_i
\varrho^{-a_i}
\eqno{(4.b)}
$$
with 
$$
\varrho=\frac{\alpha_s(m_b)}{\alpha_s(m_W)}, \ \ C_2(m_W) = -1,
\eqno{(4.c)}
$$
$$
h_i=(\frac{626126}{272277},-\frac{56281}{51730},-\frac{3}{7},-\frac{1}{14},
-0.6494,-0.0380,-0.0186,-0.0057),
$$
$$
a_i=(\frac{14}{23},\frac{16}{23},\frac{6}{23},-\frac{12}{23},0.4086,-0.4230,
-0.8994,0.1456).
\eqno{(4.d)}
$$
And $C_7(m_W) = \frac{1}{2}A(x)$, $C_8(m_W) = \frac{1}{2}C(x)$ in the 
standard model with $x=(\frac{m_t}{m_W})^2$. The functions $A ( x )$ 
and $C ( x )$ arise from graphs with $W$ boson exchange.  

In MWTCM, the relevant Feynman rules are the same as Ref.[17]:
$$
[p^+ - u_i - d_j] = i\frac{1}{\sqrt{6}F_Q}V_{u_i d_j}[m_{u_i}(1-\gamma_5) - 
m_{d_j}(1+\gamma_5)],
\eqno{(5.a)}
$$
$$
[p_8^+ - u_i - d_j] = i\frac{V_{u_i d_j}}{F_Q}\lambda^a[m_{u_i}(1-\gamma_5)
-m_{d_j}(1+\gamma_5)],
\eqno{(5.b)}
$$
where $u=( u, c, t )$, $d= ( d, s, b )$ and $V_{u_id_j}$ is the element of 
CKM matrix, and finally $F_Q$ is the decay constant of technipions composed 
of Q in MWTCM.

By explicit calculations, one can get [9]
$$
C_7(m_W)=\frac{1}{2}A(x)+\frac{1}{3\sqrt{2}G_FF_{Q}^2}[B(y)+8B(z)],     
\eqno{(5.c)}
$$
$$
C_8(m_W)=\frac{1}{2}C(x)+\frac{1}{3\sqrt{2}G_FF_{Q}^2}[D(y)
+(8D(z)+E(z))],
\eqno{(5.d)}
$$
where $y=(\frac{m_t}{m_{p^{\pm}}})^2$,\ \ $z=(\frac{m_t}{m_{p_8^{\pm}}})^2$. 
The functions $B$, $D$ and $E$ arise from diagrams with color singlet and 
color octet charged PGBs of MWTCM, and the explicit expressions for relevant 
functions are as follows:    
$$
A(x)=-\frac{x}{12(1-x)^4}[(1-x)(8x^2+5x-7)+6x(3x-2)\ln x],            
\eqno{(6.a)}
$$
$$
B(x)=\frac{x}{72(1-x)^4}[(1-x)(22x^2-53x+25)+6(3x^2-8x+4)\ln x],  
\eqno{(6.b)}
$$
$$
C(x)=-\frac{x}{4(1-x)^4}[(1-x)(x^2-5x-2)-6x\ln x],
\eqno{(6.c)}
$$
$$
D(x)=\frac{x}{24(1-x)^4}[(1-x)(5x^2-19x+20)-6(x-2)\ln x],
\eqno{(6.d)}
$$
$$
E(x)=-\frac{x}{8(1-x)^4}[(1-x)(12x^2-15x-5)+18x(x-2)\ln x].
\eqno{(6.e)}
$$

Now we write down the amplitude of Fig1.a as     
$$
\begin{array}{ll}
M_a=&\int^1_0 dx_1 dy_1 \Phi_{D_s^*}(y_1)\Phi_{B_c}(x_1)\frac{-iG_F}
{\sqrt{2}}V_{tb}V_{ ts}^*C_7(m_b)m_be\frac{\alpha_s(m_b)}{2\pi}C_F \\
&\{T_r[(\not q-m_{D_s}^{*})\not\epsilon\sigma_{\mu\nu}(1+\gamma_5)
k^{\nu}\eta^{\mu}(\not p-y_1\not q+m_b)\gamma_{\alpha}(\not p+m_{B_c})
\gamma_5\gamma^{\alpha}]\frac{1}{D_1D_3} \\
&+Tr[(\not q-m_{D_s^{*}})\not\epsilon\gamma_{\alpha}(\not q-x_1\not p)
\sigma_{\mu\nu}(1+\gamma_5)k^{\nu}\eta^{\mu}(\not p+m_{B_c})
\gamma_5\gamma^{\alpha}]\frac{1}{D_2D_3}\},    
\end{array}
\eqno{(7)} 
$$
where $\eta$ is the polarization vector of photon, $x_1$, $y_1$ are the 
momentum fractions shared by charms in $B_c$ and $D_s^*$, respectively. The 
functions $D_1$, $D_2$ and $D_3$ in equation ( 7 ) are the forms of
$$
\begin{array}{l}
D_1=(1-y_1)(m_{ B_c}^{ 2}-m_{D_s^*}^2y_1)-m_{ b}^{ 2},{\hskip 8.5cm} (8.a)\\
D_2=(1-x_1)(m_{ D_s^*}^2-m_{B_c}^2 x_1),{\hskip 9.5cm}(8.b) \\     
D_3=(x_1-y_1)(x_1m_{B_c}^{ 2}-y_1m_{D_s^*}^{ 2}).{\hskip 9cm}(8.c)
\end{array}
$$
After explicit calculation, the amplitude $M_a$ can be written as the form of
$$
M_a=i\varepsilon_{\mu\nu\alpha\beta}\eta_{\mu}k^{\nu}\epsilon^{\alpha}
p^{\beta}f_{ 1}^{ peng} +\eta^{\mu}[\epsilon_{\mu}(m_{B_c}^2-m_{D_s}^2)
-(p+q)_{\mu}(\epsilon\cdot k)]f_2^{ peng}     
\eqno{(9)}
$$
with form factors
$$
\begin{array}{ll}
f_1^{peng}&=2f_2^{peng}=C\int_{0}^{1}dx_1dy_1\delta(x_1-\epsilon_{B_c})
\delta(y_1-\epsilon_{D_s^*}) \\
&\cdot \{[m_{B_c}(1-y_1)(m_{B_c}-2m_{D_s^*}) 
-m_b(2m_{B_c}-m_{D_s^*})] {\frac{1}{D_1D_3}} - m_{B_c}m_{D_s^*}(1-x_1) 
{\frac{1}{D_2D_3}}\},
\end{array}   
\eqno{(10.a)}
$$
where
$$
C=\frac{em_bf_{B_c}f_{D_s^*}C_7(m_b)C_F\alpha_s(m_b)G_FV_{tb}
V_{ts}^*}{12\pi\sqrt{2}}.            
\eqno{(10.b)}
$$    
\begin{flushleft}
\subsection*{2.2. The weak annihilation contribution}
\end{flushleft}                                             
         
As mentioned in Sec.1, $B_c$ meson is also the unique probe of 
the weak annihilation mechanism. 

In SM, using the formalism developed by H. Y. Cheng $et \ \ al.$ [18], 
the amplitude of annihilation diagrams ( see Fig.1.b ) is
$$ 
M_{b}^{(W)}=i\varepsilon_{\mu\nu\alpha\beta}\eta^{\mu}k^{\nu}\epsilon
^{\alpha}
p^{\beta}f_{1(W)}^{anni}+\eta^{\mu}[\epsilon_{\mu}(m_{B_c}^2-m_{D_s}^2)-
(p+q)_{\mu}(\epsilon\cdot k)]f_{2(W)}^{anni}               
\eqno{(11)}
$$
with  
$$      
f_{1(W)}^{anni}=2\zeta[(\frac{e_s}{m_s}+\frac{e_c}{m_c})
\frac{m_{D_s^*}}{m_{B_c}}+(\frac{e_c}{m_c}+\frac{e_b}{m_b})]
\frac{m_{D_s^*}m_{B_c}}{m_{B_c}^2-m_{D_s^*}^2},            
\eqno{(12.a)}
$$
$$
f_{2(W)}^{anni}=-\zeta[(\frac{e_s}{m_s}-\frac{e_c}{m_c})\frac{m_{D_s^*}}
{m_{B_c}}
+(\frac{e_c}{m_c}-\frac{e_b}{m_b})]\frac{m_{D_s^*}m_{B_c}}
{m_{B_c}^2-m_{D_s^*}^2},              
\eqno{(12.b)}
$$       
where  
$$     
\zeta=ea_2\frac{G_F}{\sqrt{2}}V_{cb}V_{cs}^*f_{B_c}f_{D_s^*}, \ \ \ \ 
a_2  \ \  is \ \  a \ \  parameter.                     
\eqno{(12.c)}
$$           

In MWTCM, using the Feynman rules in equation ( 5.a ), equation ( 5.b ) 
and the methods in Ref.[18], we can write down the amplitude of charged PGBs 
annihilation diagrams ( see Fig.1.b ):  
$$
M_b^{(p)}=i\varepsilon_{\mu\nu\alpha\beta}\eta^{\mu}k^{\nu}\epsilon^
{\alpha}p^{\beta}f_{1(p)}^{anni} + \eta^{\mu}[\epsilon_{\mu}(m_{B_c}^2
-m_{D_s^*}^2)-(p+q)_{\mu}(\epsilon \cdot k)]f_{ 2(p)}^{ anni}
\eqno{(13)}
$$
with      
$$
f_{1(p)}^{anni}=-\zeta^{'}[(\frac{e_s}{m_s}+\frac{e_c}{m_c})
\frac{m_s-m_c}{m_{B_c}}+(\frac{e_b}{m_b}+\frac{e_c}{m_c})
\frac{m_b-m_c}{m_{B_c}}]\frac{m_{B_c}m_{D_s^*}}{m_{B_c}^2- m_{D_s^*}^2},      
\eqno{(14.a)}
$$
$$  
f_{2(p)}^{ anni}=\frac{1}{2}\zeta^{'}[(\frac{e_s}{m_s}+\frac{e_c}{m_c})
\frac{m_{D_s^*}}{m_{B_c}}+(\frac{e_b}{m_b}+\frac{e_c}{m_c})
]\frac{m_{D_s^{*}}m_{B_c}}{m_{ B_c}^{ 2}
-m_{ D_s^{*}}^{ 2}},        
\eqno{(14.b)}
$$                          
$$
\zeta^{'}=ea_2[\frac{2C_F}{m_{p_8^{\pm}}^2}+\frac{1}{12m_
{p^{\pm}}^2}]\frac{V_{cb}V_{cs}^*}{F_Q^2}f_{B_c}f_{D_s^*}
(m_{B_c}^2+m_{D_s^*}^2).
\eqno{(14.c)}
$$
The total annihilation amplitude ( Fig.1.b ) in the MWTCM is consequently the 
form of
$$
\begin{array}{lll}
M_b&=&M_b^{(W)} + M_b^{(p)}\\
   &=&i\varepsilon_{\mu\nu\alpha\beta}\eta^{\mu}k^{\nu}\epsilon^{\alpha}
p^{\beta}f_1^{anni} + \eta^{\mu}[\epsilon_{\mu}(m_{B_c}^2-m_{D_s^*}^2)
-(p+q)_{\mu}(\epsilon \cdot k)]f_2^{anni}
\end{array}
\eqno{(15)}
$$ 
with 
$$
f_1^{anni} = f_{1(W)}^{anni} +f_{1(p)}^{anni},
\eqno{(16.a)}
$$
$$
f_2^{anni} = f_{2(W)}^{anni} +f_{2(p)}^{anni}.
\eqno{(16.b)}
$$

\begin{flushleft}
\section*{3. Numerical results}
\end{flushleft}

We will use the following values for various quantities as input in our 
calculation.
 
(i). Decay constants for pseudoscalar $B_c$ and vector meson $D_s^*$, 
$$       
f_{D_s^*}=f_{D_s}=344 MeV   
$$
from the reports by three groups [19] and 
$$
f_{B_c}=500 MeV                                          
$$
from the results in Ref.[20].
 
(ii). Meson mass and the constituent quark mass,
$$
M_{D_s^*}=2.11GeV,  \ \ \  m_b=4.7GeV, \ \ \  m_c=1.6GeV, \ \ \  m_s=0.51GeV    
$$
from the Particle Data Group [21], and 
$$
m_{B_c}=6.27GeV                                                 
$$
as estimated in Ref.[22]. We also use $m_{B_c}\approx m_b+m_c$ , $m_{D_s^*}
\approx m_s+m_c$ in our calculation.
 
(iii). The parameter $a_2$ appearing in nonleptonic $B$ decays was 
recently extracted from the CLEO data [23] on $B\rightarrow D^* \pi(\rho)$ 
and $B\rightarrow J/\Psi K^*$ by H. Y. Cheng $et \ \ al.$ [18]. Here, we take
$$
a_2=\frac{1}{2}(c_--c_+)=0.21.
$$

(iv). For CKM elements [21], we use
$$
V_{cb}=0.04,  \ \ \  \vert V_{ts}\vert=V_{cb},  \ \ \   
\vert V_{cs}\vert=0.9745, \ \ \  V_{tb}=0.9991.  
$$

(v). The QCD coupling constant $\alpha_s(\mu)$ at any renormalization 
scale can be calculated from $\alpha_s(m_Z)=0.117$ via
$$
\alpha_s(\mu)=\frac{\alpha_s(m_Z)}{1-(11-\frac{2}{3}n_f)\frac{\alpha_s(m_Z)}
{2\pi}\ln(\frac{m_Z}{\mu})},    
$$
and we obtain 
$$
\alpha (m_b)=0.203,  \ \ \    \alpha_s(m_W)=0.119.                  
$$

(vi). For the masses of $m_{p^{\pm}}$ and $m_{p_8^{\pm}}$ in MWTCM, Ref.[12] 
has presented a constraint 
on them, here we take 
$$
m_{p^{\pm}}=(100\sim 250)GeV,                        
$$
$$
m_{p_8^{\pm}}=(300\sim 600)GeV.
$$

(vii). The decay constant $F_Q$ satisfies the following constraint [12]:
$$
F_{\pi} = \sqrt{F_{\psi}^2+3F_Q^2+N_LF_L^2}=246GeV.
$$

It is found in Ref.[12] that $F_Q=F_L=20\sim 40 GeV$. We will take
$$
F_Q=40 GeV
$$
in our calculation.

We give the long and short distance 
contributions to the form factors $f_1$ and $f_2$ in the SM and MWTCM in 
Table 1, so do the decay width in Table 2 using the amplitude formula,
$$
\Gamma(B_c\rightarrow D_s^*\gamma) = \frac{(m_{B_c}^2-m_{D_s^*}^2)^3}
{32\pi m_{B_c}^3}(f_1^2 +4f_2^2).               
$$

The lifetime of $B_c$ was given in Ref.[24]. In this paper we use
$$
\tau_{B_c}=(0.4ps\sim 1.35ps)
$$
to estimate the branching ratio $BR(B_c\rightarrow D_s^*\gamma)$ which
is a function of $\tau_{B_c}$. The results are given in Table 3. 

\begin{flushleft}
\section*{4. Discussion}
\end{flushleft}

We have studied two kinds of contributions to the process 
$B_c\rightarrow D_s^*\gamma$. For the 
short-distance one ( as illustrated in Fig.1.a ) induced by electromagnetic 
penguin,  
the momentum square of the hard scattering being exchanged by gluon is $3.6 
GeV^2$, which is large enough for PQCD analyzing. The hard scattering 
process can not be included conveniently in the soft hadronic process 
described by the wave function of the final bound state. That is one 
important reason why we 
can not apply the commonly used spectator model [25] to the two body 
$B_c$ decays. There is no phase-space for the propagators appearing in 
Fig.1.a to go on-shell, so the imaginary part of $M_a$ is absent, unlike the 
situation in Ref.[6]. Another competitive mechanism is the weak annihilation. 
In SM, we find that it is more important than the former one. This situation 
different from that of the 
radiative weak $B^{\pm}$ decays which is overwhelmingly dominated by 
electromagnetic penguin. The results stem from two reasons: one is that the 
compact size of $B_c$ meson enhances the importance of annihilation decays; 
the other comes from the Cabibbo allowance. In 
$ B_c \rightarrow D_s^* \gamma$ process, the CKM amplitude of weak 
annihilation is $\vert V_{cb}V_{cs}^*\vert $, but in $ B_{\pm} 
\rightarrow K^{\pm}\gamma$ 
process the CKM part is $ \vert V_{ub}V_{us}^* \vert $ , which is much 
smaller than $\vert V_{cb}V_{cs} \vert$.  

In addition, we find that the contribution from PGBs in MWTCM to the short 
distance process $b\rightarrow s\gamma$ is too large due to the smallness of 
the decay constant $F_Q$ in this model. In contrast, the contribution from 
PGBs through the weak annihilation process is negligibly small. In general, 
the modification from PGBs in MWTCM is too 
large to be consistent with the recent CLEO data on the branching ratio 
BR ( $b\rightarrow s\gamma$ ). 

In view of the above situation, we consider 
the TAMWTCM. The motivation of introducing topcolor to MWTCM is the 
following: in the original MWTCM, it is very difficult to generate the top 
quark mass as large as that measured in the Fermilab CDF and D0 experiments 
[26], even with `` strong " ETC [27]. Thus, topcolor interactions for the 
third generation quarks seem to be required at an energy scale of about 1 
TeV [28]. In the TAMWTCM, topcolor is still a walking theory to avoid the 
large FCNC [14]. As in other topcolor-assisted technicolor theories, the 
electroweak symmetry breaking is driven mainly by technicolor interactions 
which are strong near 1 TeV. The ETC interactions give contributions to all 
quark and lepton masses, while the large mass of the top quark is mainly 
generated by the topcolor interactions introduced to the third generation 
quarks. From Ref.[28], we can reasonably get the ETC-generated part of the 
top quark mass $m_t'=66kGeV$ with $k\sim$ 1 to  $10^{-1}$. To compare with 
the original MWTCM, we here take $m_t'=35GeV$ as the input parameter in our 
calculation. ( i.e., in the above calculations, $m_t =174GeV$ is replaced by 
$m_t'=35GeV$, the other calculations are the same as the original MWTCM ), 
The corresponding results obtained in the framework of TAMWTCM are also 
listed in the table 1, table 2 and table 3. From the results in these 
tables, we can see that the modifications from PGBs in topcolor assisted 
MWTCM to $B_c\rightarrow D_s^*\gamma$ are strongly suppressed relative to 
that in the original MWTCM. The branching ratio BR ( $B_c\rightarrow 
D_s^*\gamma$ ) in TAMWTCM is therefore consistent with the recent CLEO 
constraint on the branching ratio BR ( $b\rightarrow s\gamma$ ) for a 
certain range of the parameters.
   
In this paper, we neglected the contribution of the vector 
meson dominance VMD [29] due to the smallness of $ J/\Psi(\Psi^{'})-\gamma$ 
coupling.  
 
Finally, we estimate the possibility of observing the interesting process 
of $B_c
\rightarrow D_s^*\gamma$ at Tevatron and at the CERN Large Hadron Collider 
( LHC ). The number of 
$B_c$ at Tevatron and at LHC have estimated to be [30] 16000 ( for 25 
$Pb^{-1}$ integrated luminosities with cuts of $P_{T} ( B_c )> 10 GeV$, $y 
( B_c )<1$ ) and $2.1\times 10^8$ ( for 100 $fb^{-1}$ integrated luminosities 
with cuts of $P_T ( B_c )>20 GeV$, $y ( B_c )< 2.5$ ), respectively. 
By comparing the above predicted number of $B_c$ events with the branching 
ratio $BR_{total}^{SM}$ ( $B_c\rightarrow D_s^*\gamma$ ) as given in Table 3, 
one can understand that although this channel is unobservable at Tevatron, 
but more than one thousand events of interest will be produced at LHC, so it 
can be well studied at LHC in the future. Furthermore, it is easy to see that 
the branching ratio BR ( $B_c\rightarrow D_s^*\gamma$ ) in the TAMWTCM is 
roughly one order higher than that in the SM. Therefore, if one 
find an clear surplus of $B_c$ events in LHC experiments than that expected 
in the SM one may interpret it as a signal of new physics.

\vspace{1cm} 

\begin{flushleft}
{\bf \large  Acknowledgments}
\end{flushleft}

This work is supported by National Natural Science Foundation of China and 
the Natural Science Foundation of Henan Scientific Committee.

\newpage
\begin {center}
{\bf Reference}
\end {center}
\begin{enumerate}
\item  
B. Grinstein $et \ \ al.$, Nucl. Phys. B 339, 269 ( 1990 ).
\item 
 D. S. Du  and  Z. Wang, Phys. Rev. D 39, 1342 ( 1989 ); K. Cheng, T. C. 
Yuan, Phys. Lett. B 325, 481 ( 1994 ); Phys. Rev. D 48, 5049 ( 1994 );  
G. Lu, $et\ \ al.$, Phy. lett. B 341, 391 ( 1995 ), Phys. Rev. D 51, 2201 
(1995).
\item 
J. Tang, J. H. Liu and K. T. Chao, Phys. Rev. D 51, 3501 ( 1995 ); K. C. 
Bowler $et \ \ al.$, Phys. Rev. Lett. 72, 1398 ( 1994 ).
\item
H. Leutwyler and M. Roos, Z. Phys. C 25, 91 ( 1984 ).
\item
M. Neubert, Phys. Rep. 245, 1398 ( 1994 ).
\item 
A. Szczepaiak, $et \ \ al.$, Phys. Lett. B 243, 287 ( 1990 ); C. E. Carlson 
and J. MilRna, Phys. Lett. B 301, 237 ( 1993 ), Phys. Rev. D 49, 5908 ( 
1994 ); $ibid$ 51, 450 ( 95 ); H-n. Li and H. L. Yu, Phys. Rev. Lett 74, 
4388 ( 1995 ).
\item
Dongsheng Du, Gongru Lu, and Yadong Yang, BIHEP-TH-32 ( submitted to Phys. 
Lett.B ).
\item 
S. J. Brodsky and G. P. Lepage, Phys. Rev. D 22, 2157 ( 1980 ).
\item
Cai-Dian L\"u and Zhenjun Xiao, Phys. Rev. D 53, 2529 ( 1996 ).
\item 
S. Dimopoulos and L. Susskind, Nucl. Phys. B 155, 237 ( 1979 ); E. Eichten 
and K. Lane, Phys. Lett. B 90, 125 ( 1980 ).
\item
T. Appelquist and G. Trintaphyllon, Phys. Lett. B 278, 345 ( 1992 ); R. 
Sundrum and S. Hsu, Nucl. Phys. B 391, 127 ( 1993 ); T. Appelquist and J. 
Terning, Phys. Lett. B 315, 139 ( 1993 ). 
\item 
K. Lane and E. Eichten, Phys. Lett. B 222, 274 ( 1989 ); K. Lane and M. V. 
Ramana, Phys. Rev. D 44, 2678 ( 1991 ).
\item
Chongxing Yue, Yuping Kuang and Gongru Lu, TUIMP-TH-95/72.
\item
C. T. Hill, Phys. Lett. B 345, 483 ( 1995 ); K. Lane and E. Eichten, Phys. 
Lett. B 352, 382 ( 1995 ). 
\item
M. Battle $et \ \ al.$, CLEO Collaboration, Preprint, CLEO 93-13.
\item
S. J. Brodsky and C. R. Ji, Phys. Rev. Lett 55, 2257 ( 1985 ).
\item 
Zhenjun Xiao, Lingde Wan, Jinmin Yang and Gongru Lu, 
Phys. Rev. D 49, 5949 ( 1994 ).
\item 
H. Y. Cheng, $et \ \ al.$, Phys. Rev. D 51, 1199 ( 1995 ).
\item 
A. Aoki, $et \ \ al.$, Prog. Theor. Phys. 89, 137 ( 1993 ); D. Acosta 
$et \ \ al.$, CLNS 93/1238; J. Z. Bai, $et \ \ al.$, BES Collaboration Phys. 
Rev. Lett. 74, 4599 ( 1995 ).
\item 
W. Buchm\"uiller and S-H. HTye, Phys. Rev. D 24, 132 ( 1994 ); A. Martin, 
Phys. Lett. B 93, 338 ( 1980 ); C. Quigg and J. L. Rosner, Phys. Lett. B 71, 
153 ( 1977 ); E. Eichten $et \ \ al.$, Phys. Rev. D 17, 3090 ( 1978 ). 
\item 
Particle Data Group, 
L. Montanet $et \ \ al.$, Phys. Rev. D 50, 1173 ( 1994 ).
\item 
W. Kwong and J. L. Rosner, Phys. Rev. D 44, 212 ( 1991 ).
\item 
CLEO Collaboration, M. S. Alam $et \ \ al.$; Phys. Rev. D 50, 43 ( 1994 ).
\item 
C. Quigg, FERMILAB-Conf-93/265-T; C. H. Chang and Y. Q. Chen, Phys. Rev. 
D 49, 3399 ( 1994 ); P. Colangelo, $et \ \ al.$, Z. Phys. C 57, 43 ( 1993 ). 
\item 
M. Bauer $et \ \ al.$, Z. Phys. C 29, 637 ( 1985 ); B. Grinstein 
$et \ \ al.$, Phys. Rev. D 39, 799 ( 1987 ).
\item
The CDF Collaboration, F.Abe et al., Phys. Rev. Lett. 74, 2626 ( 1995 ); 
The D0 Collaboration, S.Abachi et al., Phys. Rev. Lett. 74, 2697 ( 1995 ).
\item
T. Appelquist, M. B. Einhorn, T. Takeuchi and L. C. R. Wijewardhana, Phys. 
Lett. B 220, 223 ( 1989 ); R. S. Chivukula, A. G. Cohen and K. Lane, Nucl. 
Phys. B 343, 554 ( 1990 ); A. Manohar and H. Georgi, Nucl. Phys. B 234, 189 
( 1984 ).
\item
C. T. Hill, Phys. Lett. B 266, 419 ( 1991 ); S. P. Martin, Phys. Rev. D 45, 
4283 ( 1992 ); D 46, 2197 ( 1992 ); Nucl. Phys. B 398, 359 ( 1993 ); M. 
Linder and D. Ross, Nucl. Phys. B 370, 30 ( 1992 ); C. T. Hill, D. Kennedy, 
T. Onogi and H. L. Yu, Phys. Rev. D 47, 2940 ( 1993 ); W. A. Bardeen, C. T. 
Hill and M. Lindner, Phys. Rev. D 41, 1649 ( 1990 ).  
\item 
E. Golowich and S. Pakvasa, Phys. Rev. D 51, 1215 ( 1995 ); H. Y. Cheng, 
Phys. Rev. D 51, 6288 ( 1995 ).
\item 
K. Cheung, Phys. Rev. Lett 71, 3413 ( 1993 ). 
\end{enumerate}

\newpage
\begin{table}[h]
 \caption{Form factors in the SM, MWTCM and TAMWTCM. $f^{peng}$ and 
$f^{anni}$ represent form factors for electromagnetic penguin and weak 
annihilation process, respectively. }
 \begin{center}
 \begin{tabular}{|c|c|c|c|} \hline
 $f_i$ & SM & MWTCM & TAMWTCM \\ \hline
 $f_1^{peng}$ &$ -3.05\times 10^{-10}$ & $(0.50 \sim 1.13)\times 10^{-8}$ 
& $(0.44\sim 1.67)\times 10^{-9}$   \\ \hline
$f_2^{peng}$ & $-1.57\times 10^{-10}$ &$(2.50\sim 5.65)\times 10^{-9}$ 
&$(2.22\sim 8.38)\times 10^{-10}$  \\ \hline
$f_1^{anni}$ & $7.10\times 10^{-10}$ & $(6.75\sim 7.02)\times10^{-10}$ 
&$(6.75\sim 7.02)\times 10^{-10}$   \\ \hline
$f_2^{anni}$ & $-1.70\times 10^{-10}$ & $(-1.66\sim -1.53)\times 10^{-10}$ 
&$(-1.66\sim -1.53)\times 10^{-10}$\\ \hline    
\end{tabular}
 \end{center}
 \end {table}

\begin{table}[h]
\caption{The decay rates in the SM, MWTCM and TAMWTCM. The $\Gamma^{peng}$, 
$\Gamma^{anni}$ and $\Gamma^{total}$ represent $\Gamma ( B_c\rightarrow 
D_s^*\gamma )$ through penguin, annihilation, and penguin + annihilation 
diagrams, respectively. }
 \begin{center}
 \begin{tabular}{|c|c|c|c|}  \hline
 $\Gamma(B_c \rightarrow D_s^* \gamma)$  & SM  & MWTCM & TAMWTCM \\ \hline
 $\Gamma^{peng}( GeV )$  &$ 3.18 \times 10^{-19}$ & $(0.86\sim 4.36)
\times 10^{-16}$ & $(0.67\sim 9.55)\times 10^{-18}$ \\ \hline
 $\Gamma^{anni}(GeV)$  & $1.06\times 10^{-18}$ & $(0.94\sim 1.03)\times 
10^{-18}$ &$(0.94\sim 1.03)\times 10 ^{-18}$ \\ \hline
$\Gamma^{total}(GeV)$  & $9.92\times 10^{-19}$ & $(0.93\sim 4.52)\times 
10^{-16}$ &$(0.23\sim 1.26)\times 10^{-17}$ \\ \hline
 \end{tabular}
 \end{center}
 \end{table}

\begin{table}[h]
\caption{The branching ratios ( $B_c\rightarrow D_s^*\gamma$ ). The 
$BR_{total}^{SM}$, $BR_{total}^{MWTCM}$, $BR_{total}^{TAMWTCM}$ represent 
the branching ratio ( $B_c\rightarrow D_s^*\gamma$ ) in the SM, MWTCM and 
TAMWTCM, respectively. }
\begin{center}
\begin{tabular}{|c|c|c|c|} \hline
$\tau_{B_c} $ & 0.4ps & 1.0ps & 1.35ps  \\ \hline
$ BR^{SM}_{total} $ & $6.03\times 10^{-7}$ & $
1.51\times 10^{-6}$ & $2.04\times 10^{-6}$ \\ \hline
$BR^{MWTCM}_{total}$ &$(0.57\sim 2.75)\times 10^{-4}$ 
&$(1.42\sim 6.86)\times 10^{-4}$ &$(1.91\sim 9.26)
\times 10^{-4}$ \\ \hline
$ BR^{TAMWTCM}_{total}$ &$(1.37\sim 7.66)\times 10^{-6}$ &$(0.34\sim 1.91)
\times 10^{-5}$ &$(0.46\sim 2.58)\times 10^{-5}$\\ \hline  
\end{tabular}
\end{center}
\end{table} 

\vspace{1cm}
\begin{center}
Figure caption
\end{center}

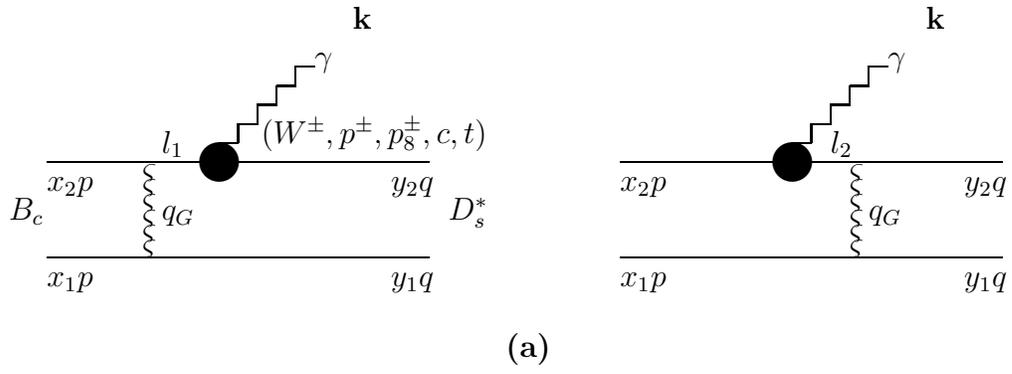
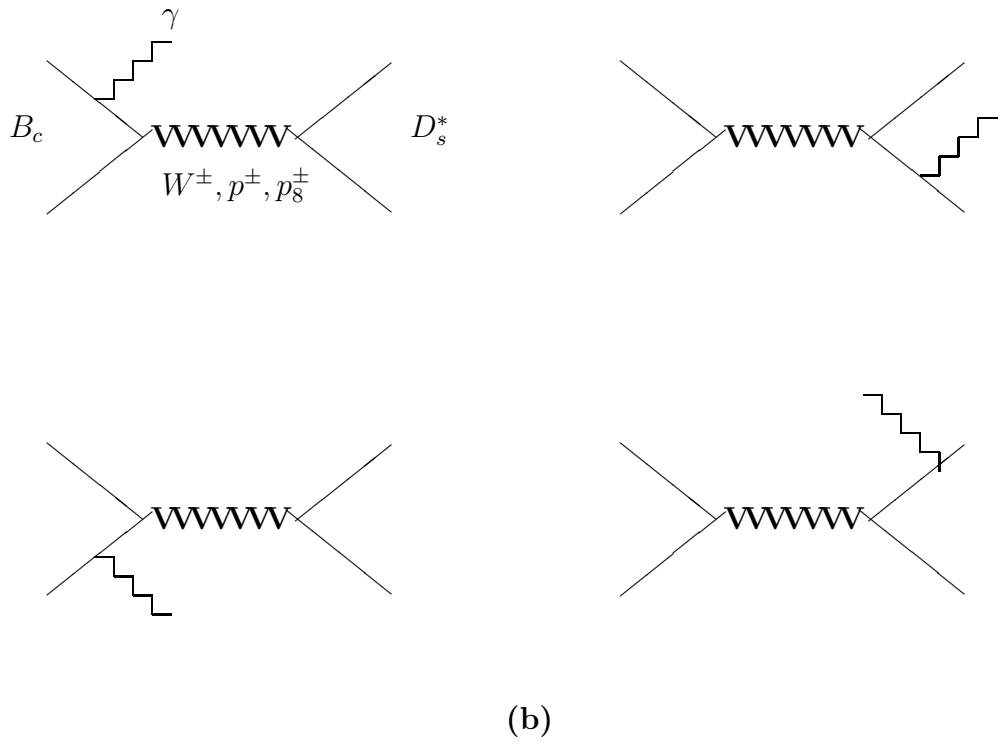
Fig.1: Fig.1.a shows the Feynman diagrams which contribute to the decay 
$B_c\rightarrow D_s^*\gamma$ through the short distance $b\rightarrow 
s\gamma$ mechanism. The blob represents the electromagnetic penguin 
operators contributing to $b\rightarrow s\gamma$. $x_2p$ and $x_1p$ are 
momenta of $b$ and $c$ quarks in the $B_c$ meson, respectively. $y_2q$ and 
$y_1q$ are momenta of $s$ and $c$ quarks in the $D_s^*$ meson, respectively. 
Fig.1.b represents the Feynman diagrams which contribute to the decay 
$B_c\rightarrow D_s^*\gamma$ through the weak annihilation.

\newpage

\begin{picture}(30,0)
{\bf 
\setlength{\unitlength}{0.1in}
\put(5,-15){\line(1,0){20}}
\put(5,-20){\line(1,0){20}}
 \put(3,-18){$B_c$}  
  \put(26,-18){$D_s^*$}
   \multiput(10,-15.8)(0,-0.8){6}{$\varsigma$}
   \multiput(14,-15)(1,1){5}{\line(0,1){1}} 
   \multiput(13,-15)(1,1){6}{\line(1,0){1}}
    \put(14,-15){\circle*{3}}
      \put(11,-18){$q_G$}
\put(16.2,-14){$(W^{\pm}, p^{\pm}, p_8^{\pm}, c, t)$}
\put(5,-16.5){$x_2 p$}
\put(5,-21.5){$x_1 p$}
\put(23,-16.5){$y_2 q$}
\put(23,-21.5){$y_1 q$}
\put(11,-14.5){$l_1 $}
\put(29,-25.2){(a)}
\put(21,-8){k}
\put(19,-10){$\gamma$} 
\put(35,-15){\line(1,0){20}}
\put(35,-20){\line(1,0){20}}
   \multiput(47,-15.8)(0,-0.8){6}{$\varsigma$}
   \multiput(44,-15)(1,1){5}{\line(0,1){1}} 
   \multiput(43,-15)(1,1){6}{\line(1,0){1}}
\put(48,-18){$q_G$}
\put(35,-16.5){$x_2 p$}
\put(35,-21.5){$x_1 p$}
\put(53,-16.5){$y_2 q$}
\put(53,-21.5){$y_1 q$}
\put(46,-14.5){$l_2 $}
\put(44,-15){\circle*{3}} 
\put(51,-8){k}
\put(49,-10){$\gamma$}
\put(5,-35){\line(5,-4){5}}
\put(5,-43){\line(5,4){5.5}}
\multiput(10.4,-39.5)(1,0){7}{V}
\put(17.5,-38.5){\line(5,-4){5.5}}
\put(18,-39.1){\line(5,4){5}}
   \multiput(51.7,-41)(1,1){3}{\line(0,1){1}} 
   \multiput(50.7,-41)(1,1){4}{\line(1,0){1}}
 \put(3,-39){$B_c$}  
  \put(24,-39){$D_s^*$}
\put(11,-42){$W^{\pm}, p^{\pm}, p_8^{\pm} $}
\put(11,-33){$\gamma$}
\put(35,-35){\line(5,-4){5}}
\put(35,-63){\line(5,4){5.5}}
\multiput(40.4,-59.5)(1,0){7}{V}
\put(47.5,-58.5){\line(5,-4){5.5}}
\put(48,-59.1){\line(5,4){5}}
\multiput(8.5,-61)(1,-1){3}{\line(0,-1){1}} 
\multiput(7.5,-61)(1,-1){4}{\line(1,0){1}}
\put(5,-55){\line(5,-4){5}}
\put(5,-63){\line(5,4){5.5}}
\multiput(10.4,-59.5)(1,0){7}{V}
\put(17.5,-58.5){\line(5,-4){5.5}}
\put(18,-59.1){\line(5,4){5}}
   \multiput(51.7,-55.5)(-1,1){4}{\line(0,-1){1}} 
   \multiput(50.7,-55.5)(-1,1){4}{\line(1,0){1}}
\put(35,-55){\line(5,-4){5}}
\put(35,-43){\line(5,4){5.5}}
\multiput(40.4,-39.5)(1,0){7}{V}
\put(47.5,-38.5){\line(5,-4){5.5}}
\put(48,-39.1){\line(5,4){5}}
\multiput(8.5,-37)(1,1){3}{\line(0,1){1}} 
\multiput(7.5,-37)(1,1){4}{\line(1,0){1}}
\put(29,-70){(b)} 
\put(29,-80){Fig.1} 
}
\
\end{picture}
\end{document}